\documentclass{epl}
\usepackage{latexsym,amssymb,amsmath}

\newcommand{\ie}{{\it i.e.\/}}

\title{Particle Survival and Polydispersity in Aggregation}
\shorttitle{Particle survival ...}
\author{E. K. O. Hell\'en, P. E. Salmi \and M. J. Alava}

\shortauthor{E. Hell\'en}
\institute{Laboratory of Physics, Helsinki University of Technology,
P. O. Box 1100, FIN-02150 HUT, Finland}

{Fluctuation phenomena, random processes, noise, and Brownian motion}

\begin{document}
\newcommand{\figwidth}{0.5\textwidth}
\newcommand{\figwidthdual}{0.4\textwidth}

\maketitle
\begin{abstract}
We study the probability, $P_S(t)$, of a cluster to remain intact in
one-dimensional cluster-cluster aggregation when the cluster diffusion
coefficient scales with size as $D(s) \sim s^\gamma$. $P_S(t)$
exhibits a stretched exponential decay for $\gamma < 0$ and the
power-laws $t^{-3/2}$ for $\gamma=0$, and $t^{-2/(2-\gamma)}$ for
$0<\gamma<2$. A random walk picture explains the discontinuous and
non-monotonic behavior of the exponent. The decay of $P_S(t)$
determines the polydispersity exponent, $\tau$, which describes the
size distribution for small clusters. Surprisingly, $\tau(\gamma)$ is
a constant $\tau = 0$ for $0<\gamma<2$. 

\end{abstract}

Many models of aggregation phenomena lead to scale-invariance:
the average cluster size increases as a power-law, $S(t) \sim t^z$,
which defines a dynamical exponent $z$.  
This kind of behavior is met in various contexts ranging from
chemical engineering to materials science to atmosphere research to,
ultimately, even astrophysics~\cite{Family_Landau}. 
It is of interest to explore the statistics of aggregation as a dynamical
process, beyond the length- and timescales defined through $z$. 

In this Letter we introduce a new quantity in aggregation systems, the
cluster survival, defined as the probability $P_S(t)$, that a cluster 
present at $t=0$ remains unaggregated until time $t$. This is a first passage
problem~\cite{rednerinkirja} in a many body system and analogous to
persistence which is often studied by measuring the fraction of
a system that preserves its initial condition for all times $[0,t]$
\cite{Majumdar:CurrSciIndia}. The cluster survival turns out to decay
in a nontrivial and counterintuitive manner. The behavior can be
understood by a mean-field like random walk analysis. However, even on
the mean-field 
level the question reduces to a novel, unsolved random walk problem,
which we analyse in the long time limit. More importantly, by
solving the decay of the cluster survival we are able to determine
the polydispersity exponent characterising the cluster size
distribution.

We concentrate on a common and important
example: diffusion--limited
cluster--cluster aggregation (DLCA)~\cite{Meakin:PhysicaScripta46}.
In the lattice version of DLCA any set of
nearest neighbor occupied lattice sites is identified as a cluster.
Each of these performs a random walk with a size dependent diffusion
constant, $D(s) \sim s^\gamma$, where $\gamma$ is the diffusion exponent.
Colliding clusters are merged together and the aggregate
diffuses either faster ($\gamma>0$) or slower than before ($\gamma<0$).
In the following cluster survival is investigated in the one-dimensional
case for numerical and analytical simplicity. We employ numerical
simulations and random walk~(RW) arguments.

The numerics is made transparent
by mapping the behavior of surviving clusters to a three-particle
RW picture: two particles with a time-dependent diffusion
coefficient confine a surviving one
which diffuses at a constant rate. This is an analogy of the famous
independent interval approximation often used in persistence studies
\cite{MajumdarDerrida:PRL77}.  One can discern three separate
cases: first, 
$0 < \gamma < 2$, which results in a power-law decay for the survival:
$P_S(t) \sim t^{-\theta_S(\gamma)}$, second, $\gamma=0$ which is
exactly solvable both for the RW and the DLCA problems, and $\gamma <0$ 
when $P_S$ decays as a stretched exponential. For $\gamma >2$ the
system has a gelation transition and is not of interest here.

The RW problem has, to our knowledge, not been discussed in the
literature. We
consider the asymptotic behavior of the associated Fokker-Planck
equation and perform numerical simulations.
The first main result is that $\theta_S (\gamma)  = 2/(2-\gamma)$
when $0 < \gamma < 2$.
Thus the survival is {\em discontinuous} and
{\em non-monotonic} since $\theta_S(0) = 3/2$
but $\theta_S(0^+) = 1$.
This non-intuitive result follows since for $\gamma >0$
the RW problem becomes asymptotically separable as the ratio of the
diffusion coefficients diverges: the surviving clusters are immobile.
For $\gamma < 0$ the same is not true and the fluctuations of
the confining clusters remain relevant in determining the
stretching exponent.

One of the main interests in aggregation is the behavior of the cluster
size distribution, $n_s(t)$ (the number of cluster of size $s$ 
per lattice site at time $t$). 
For DLCA simulations and experiments have validated the
scaling~\cite{Meakin:PhysicaScripta46}
$
n_s(t) = S(t)^{-2} f\left(s/S(t)\right), \label{nstscaleq}
$
where
the scaling limit, $s \to \infty$, $S(t) \to  \infty$
with $x \equiv s/S(t)$ fixed, is
taken. In one dimension
$z = 1/(2-\gamma)$~\cite{KangMiyazima:PRA33}. 

There is a fundamental difference between $\gamma<0$ and $\gamma
\ge 0$. For $\gamma<0$ the cluster size distribution is bell-shaped
and $f(x)$ decays faster than any power at both tails. 
For $\gamma \ge 0$ it is broad so that $f(x) \sim x^{-\tau}$
($x \to 0$)
defines the polydispersity exponent, $\tau$, which characterizes the 
density of small clusters.
In this region we show evidence for the scaling relation $\theta_S =
(2-\tau)z$. Therefore $\tau$ is determined by the cluster survival
strategy, shedding light on the non-trivial problem how
to compute it~\cite{vanDongen:PRL54,Cueille:PRE55}. The final main
result follows thus: 
$\tau(\gamma) = 0$ for $0<\gamma<2$ indicating a flat cluster size
distribution. This is confirmed by simulations.

Next we present a mean-field random walk analysis to calculate
the survival exponent $\theta_S(\gamma)$. Although the kinetics in
one-dimension is fluctuation-dominated~\cite{Kang:PRA30},
the mean-field approach turns out to capture the essential ingredients
for $\gamma\ge0$. This is demonstrated by comparing the random walk
survival to that of the full DLCA one. In the opposite case,
when $\gamma < 0$, the stretching exponent obtained from
RW simulations differs from the DLCA case.
In the limit $t \gg 0$ the average distance between
clusters grows as $t^{z}$ and between the surviving
ones as $t^{\theta_S}$. The latter become
separated by aggregated clusters at late times, since evidently
$\theta_S > z$. 
Thus it is sufficient to consider only one trial cluster
and its two neighbors. These
grow still by collisions with neighbors on the opposite side.
In the mean-field approximation the discrete growth events can be substituted
by a continuous process and the neighboring clusters grow
as the average cluster does. 
The finite extent of clusters is irrelevant and they
can be considered as point particles. Let
$x_i(t)$ ($i=1,2,3$) denote their positions
at time $t$ with $x_1(0) < x_2(0) < x_3(0)$.

The motion of these particles is described by the
Langevin equations
\begin{equation}
\dot{x}_i(t) =  \xi_i(t) \label{Lang1}
\end{equation}
with Gaussian white noises $\langle \xi_i(t) \rangle
= 0$ and $\langle \xi_i(t) \xi_j(t') \rangle =  2 {\cal D}_i(t)
\delta_{ij} \delta(t-t')$, in the standard notation.
The diffusion coefficients read as ${\cal D}_1(t)
= {\cal D}_3(t) = D_1 t^{\gamma z}$ and ${\cal D}_2(t) = D_2$. The
time dependent diffusion constant, say ${\cal D}_1(t)$, implies
that the particle $1$ will follow a simple diffusive motion with a
constant diffusion coefficient $D_1$ in a time scale
$
T(t) = \int_0^t {\rm d}t' {\cal D}_1(t')/D_1 = t^{\gamma z +
1}/(\gamma z + 1). 
$

The survival of the center particle is determined by the
termination of the process, given by either $x_1(t) =
x_2(t)$ or $x_2(t)=x_3(t)$. It is natural to consider
the distances between the particles: $x_{12}(t) = x_2(t) - x_1(t) \ge
0$ and $x_{23}(t) = x_3(t) - x_2(t) \ge 0$.
Starting from Eq.~\ref{Lang1} the following
Fokker-Planck equation is reached:
\begin{equation}
\frac{\partial{\rho}}{\partial t} = (D_2+D_1t^{\gamma z})
\left(
 \frac{\partial^2{\rho}}{\partial x_{12}^2}
 + \frac{\partial^2{\rho}}{\partial x_{23}^2}
\right)
- 2 D_2 \frac{\partial^2{\rho}}{\partial x_{12} \partial x_{23}},
\label{FPeq}
\end{equation}
where $\rho(x_{12},x_{23};t)$ is the probability density for the
two distances at time $t$. The initial condition is now
$\rho(x_{12},x_{23};0) = \delta(x_{12}-x_{12}(0))
\delta(x_{23}-x_{23}(0))$.
The termination of the process when two particles collide gives
absorbing boundary conditions along the axis, \ie,
$\rho(x_{12},0;t)=0$ and $\rho(0,x_{23};t) = 0$
for all times $t$.
The survival probability
\begin{equation}
P_{\rm RW}(t) = \int_0^\infty {\rm d}x_{12} \int_0^\infty
{\rm d}x_{23}\ \rho(x_{12},x_{23};t) \sim t^{-\theta_{RW}(\gamma)},
\end{equation}
where the exponent $\theta_{RW}$ is the survival exponent for the RW
problem. 

Equation~\ref{FPeq} can not be solved for $t$ arbitrary since
the absorbing boundary conditions together with the two different
time scales make the standard methods inappropriate.
However, the survival exponent is given by the leading-order
asymptotic behavior when $t \to \infty$. We consider the
large time limit and the three different cases separately:
$\gamma < 0$, $\gamma = 0$, and $\gamma > 0$.
In the size independent case, $\gamma = 0$, 
the collisions of the clusters surrounding a surviving cluster
with other clusters do not matter. This is
an old problem of three similar
annihilating random walkers, for which the survival exponent is known
to be $\theta_{RW}(0) = 3/2$~\cite{Fisher:JSP34}. Also the full DLCA
problem can be solved exactly with the result $\theta_S(0) =
3/2$\cite{Spouge:PRL60}.

Diagonalizing Eq.~\ref{FPeq} in the time scale $T$
gives a simple diffusion equation in a wedge with two absorbing 
boundaries. However, for $\gamma \neq 0$ the wedge angle is a function
of time making the exact solution hard.
For $\gamma >0$ the leading term of the survival probability at late
times can be 
obtained by considering the survival in the final wedge angle
$\Theta_\infty 
= \pi/2$. Thus~\cite{rednerinkirja} $P_{\rm RW}(t) \sim
T^{-\pi/2\Theta_\infty} \sim 
T^{-1} \sim t^{-(1+\gamma
z)}$. Since $z=1/(2-\gamma)$ the survival exponent reads
$\theta_{RW}(\gamma) = 2z=2/(2-\gamma)$.
The approximation obtained by replacing
the time-dependent angle by the final opening angle
corresponds to putting $D_2=0$ in
Eq.~\ref{FPeq}, \ie, to taking the center particle
to be at rest. This guess can be validated by
directly solving equation~\ref{FPeq} and analyzing the limiting
behavior of the solution~\cite{omapitka}.
Note, that in this limit $\theta_{RW}(\gamma)$ can be simply determined
from two independent random walkers with a {\it fixed} absorbing
boundary in between.

For $\gamma<0$ the situation is more tricky and will be considered in
more detail elsewhere~\cite{omapitka}. Briefly, proceeding similarly as
above leads to a closing wedge with final angle $\Theta_\infty =
0$. This would correspond to a
situation where now the particles $1$ and $3$ are fixed and therefore
to a simple exponential decay for the survival~\cite{rednerinkirja}.  
However, simulations show that the expectation value of their positions
grows like $t^\alpha$ with non-trivial $\alpha$.
The survival becomes
a stretched exponential $P_{\rm RW}(t) \sim \exp(-C t^{\beta_{\rm
RW}})$ with $C>0$ a constant. 
If the distance
between the outermost particles would grow {\em deterministically\/} as
$t^\alpha$ this would lead to a stretched exponential decay with
$\beta_{\rm det} = 1-2\alpha$~\cite{rednerinkirja}. Here the
fluctuations in the particle 
position violate this relation.

The RW picture is tested by comparing it to DLCA by simulations.
These are done in the usual fashion,
with a monodisperse initial condition and equal distances
between neighboring clusters.
$P_S$ is independent of the initial distribution
except for transient effects~\cite{omapers}.

In the case $\gamma = 0$ with the initial distances
being $x_{12}(0)=x_{23}(0)=l_0$ the first correction to scaling
reads~\cite{Derrida:PRE54}
$
P_S(t) =
\frac{1}{4\sqrt{2\pi}} \left( \frac{l_0^2}{Dt}\right)^{3/2} \times \left[
1-\frac{3}{16}\left(\frac{l_0^2}{Dt}\right) + {\cal O} \left(
\left(\frac{l_0^2}{Dt}\right)^{-2} \right) \right]
$
so that the correction becomes negligible for times much larger than
the cross-over time
$t_{\rm cr} = 3l_0^2/(16Dt)$.
For $\gamma > 0$ the ratio of the diffusion coefficients, $r=
D_2/D_1t^{\gamma z}$, controls the validity of the approximation of
neglecting the constant terms in Eq.~\ref{FPeq}. Therefore the
cross-over time depends on $\gamma$ as $t_{\rm cr} \sim
r^{(2-\gamma)/\gamma}$, which diverges for $\gamma \to 0$. We can thus
expect that the asymptotic scaling regime can be reached in
simulations only for relatively large values of $\gamma$.

\begin{figure}
\centering
\includegraphics[angle=-90,width=\linewidth
]{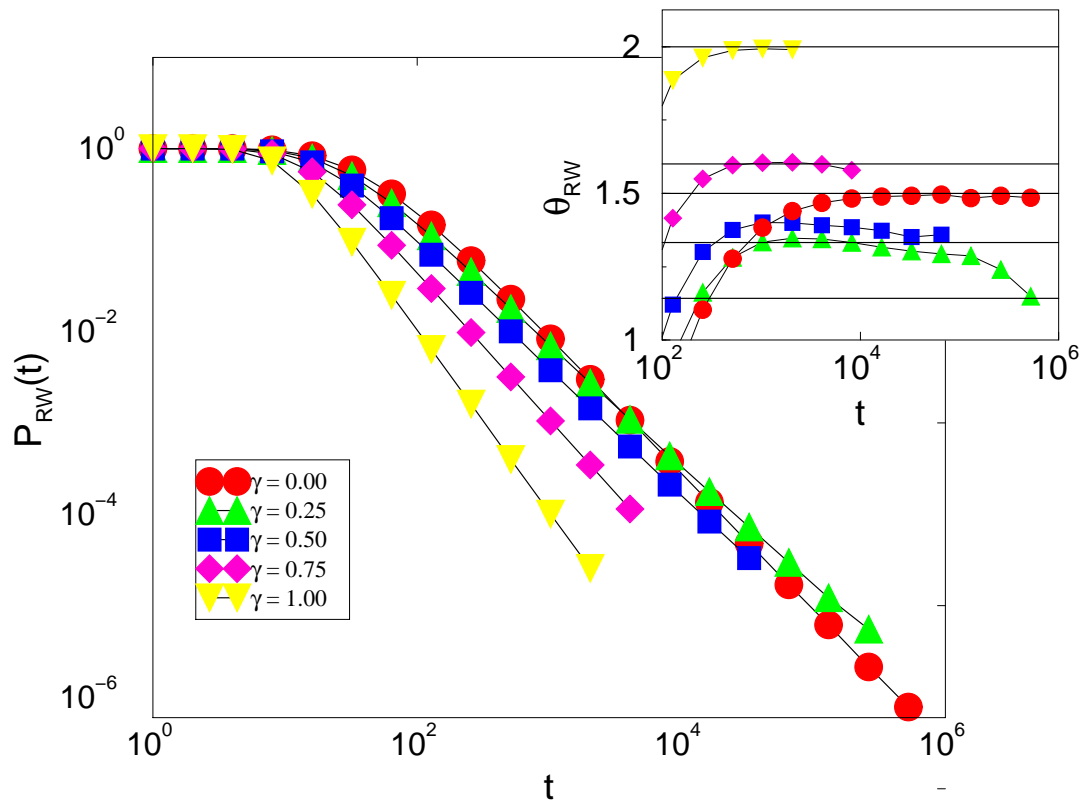}
\caption{The survival probability of three random walkers.
The inset shows the corresponding local exponents. The
solid lines correspond to the analytic values given
by $\theta_{RW}= 2/(2-\gamma)$. The data are averaged over
variable number of realizations ranging from $10^9$ for $\gamma = 0$
to $2\times10^7$ for $\gamma = 0.5$. The initial distance between particles is
$10$.
}
\label{surv1fig}
\end{figure}

Figure~\ref{surv1fig} shows the survival probabilities obtained from
simulations. $P_{\rm RW}(t)$ clearly decays as a power-law for
large times. For $\gamma = 0$ the survival exponent  saturates to
the asymptotic value $\theta_{RW} = 3/2$ around $t \approx 10^4$ as
shown in the inset, where the local survival exponent
is presented.
The exponent saturates also for $\gamma = 0.75$ and
$\gamma=1$. For $0<\gamma<0.75$ the survival exponent is slowly
approaching the asymptotic value given by $\theta_{RW} = 2/(2-\gamma)$.
For $\gamma = 0.75$ $r \approx 60$ at the time when the exponent
saturates. This would correspond to $t_{\rm cr} \approx 2 \times 10^5$
and $3 \times 10^{12}$ for
$\gamma = 0.50$ and $0.25$, respectively. Thus we can not reach the
asymptotic regime for $\gamma \lesssim 0.5$.

\begin{figure}
\centering
\includegraphics[angle=-90,width=\linewidth
]{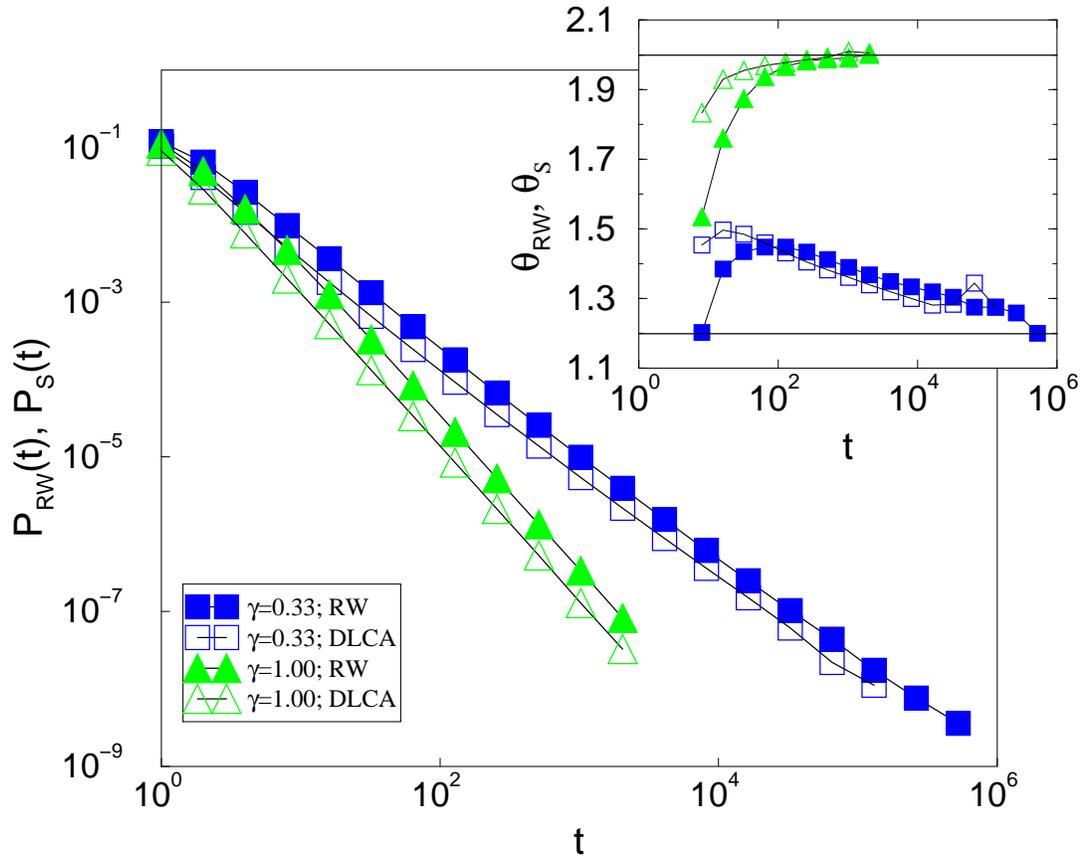}
\caption{a) Comparison between the random walk 
and DLCA survival probabilities.
 b)
The corresponding local exponents.  The
solid lines correspond to the analytic values given by
$2/(2-\gamma)$. The RW simulations are averaged over
$2.5 \times 10^{10}$ ($\gamma = 0.33$) and $2\times10^{10}$ ($\gamma =
1.00$)
realizations with the initial distance between particles being 2. The
DLCA simulations are averaged over $50000$ realizations on a
system of size $55555$.
}
\label{pers1fig}
\end{figure}

In Figure~\ref{pers1fig} the DLCA and RW survival probabilities are 
compared. The two behaviors agree except for a small difference
between the amplitudes. The initial inter-particle  
distances are taken to be the same, in order the RW-picture
to be as close to DLCA as possible. 
Notice the figure contains a case ($\gamma =0.33$) in which
the asymptotic regime is not reached.

For $\gamma<0$ the survival probability decays stretched
exponentially for both RW and DLCA systems (not shown).  However, the 
stretching exponents differ from each other. This further supports the
importance of fluctuations for $\gamma<0$.  For RW survival we obtain 
$\beta_{\rm RW} \approx 0.33,0.31,0.25,$ and 0.19 for $\gamma =
-6,-4,-2,$ and $-1$, respectively. 
For DLCA the numerics suggest an expression $\beta_S = 2/3(1-2z)$. 
We also separately checked that the average size of the neighbors of
unaggregated clusters in DLCA grows as $t^z$.
Note, that $\lim_{\gamma \to -\infty} \beta_S =
2/3$ and not a simple exponential ($\beta_S = 1$) 
as one might expect based on the case of 
two immobile neighbors. 
It is also quite surprising, that the mean field approximation
  works better for a broad cluster size distribution
  ($\gamma \ge  0:f(x)\sim x^{-\tau} (x\to0)$) than in the case, where
  the distribution is narrow around the mean ($\gamma < 0:f(x)$ decays
  faster than any power law at both ends).

To summarize the results the survival probability decays as
\begin{equation}
P_S(t)  \sim
\begin{cases}
\exp(- C t^{\beta_S})  &, \gamma < 0 \\
t^{-3/2}  &, \gamma = 0 \\
t^{-2/(2-\gamma)}  &, 0<\gamma < 2. \label{varthetagamma}
\end{cases}
\end{equation}
The survival exponent is discontinuous
at $\gamma = 0$, \ie, $3/2 = \theta_S(0) > \theta_S(0^+) =
1$. This seems first counterintuitive since making some of the
particles to diffuse faster helps the other to survive longer!
This is the best since the surviving particle
eventually ``discovers'' the optimal strategy~\cite{Redner:AJP67} of
remaining 
stationary. This is further confirmed by the fact that the probability
of finding a site 
that has never been under any of the clusters decays as power-law
with the exponent $2/(2-\gamma)$ for all $\gamma <
2$~\cite{omapers}. For $\gamma<0$ the surviving
clusters no longer remain stationary.

The third exponent of interest in DLCA scaling
is the decay exponent, $w$, which
describes the decrease of the number of clusters of a fixed size
$s$ as a function of time $n_s(t) \sim t^{-w}$. As the other
exponents $z$ and $\tau$, it is expected to depend on $\gamma$.
The three exponents are related by
the scaling relation~\cite{Vicsek:PRL52}
$
w = (2-\tau)z.
$
Therefore dynamic scaling is fully characterized by any two
of the exponents. Even on mean-field level (Smoluchowski's equation)
the only easy exponent is $z$ since it does
not involve the full scaling function as in the case of  and
$\tau$ and $w$\cite{vanDongen:PRL54}.

For $\gamma = 0$ an exact solution of the cluster size distribution
$n_s(t)$ is  possible,
 with $w=3/2$ for any short-range correlated
initial distribution $n_s(0)$~\cite{Spouge:PRL60}. For a monodisperse
initial condition, 
$n_s(0) = \delta_{1,s}$, the survival probability is simply $n_1(t)$,
yielding the exponent $\theta_S(0) = 3/2$.
For $\gamma>0$ the dynamics of the clusters in the $s \ll S(t)$ part
of the size distribution is dictated by collisions
with larger, faster ones, that remove such small clusters
from the tail. The mechanism by which clusters stay in the tail 
should be the same as for the survival problem. Thus 
for $0\leq \gamma<2$ one should have for the decay
exponent $w = \theta_S= 2/(2-\gamma)=2z$ together with the scaling relation
\begin{equation}
\theta_S = (2-\tau)z. \label{scalrelC}
\end{equation}
The numerically
estimated values for the exponents fulfill Eq.~\ref{scalrelC}
within the error bars for all values of the diffusion
exponent $\gamma$~\cite{omapitka}.
Hence, the
polydispersity exponent is discontinuous at $\gamma=0$ since
$\tau(0) = -1  \neq  0 = \tau(0^+)$, similarly to some examples
on the mean-field level~\cite{vanDongen:PRA32}. It is also surprisingly enough
independent of the value of~$\gamma$.

\begin{figure}
\centering
\includegraphics[width=\linewidth
]{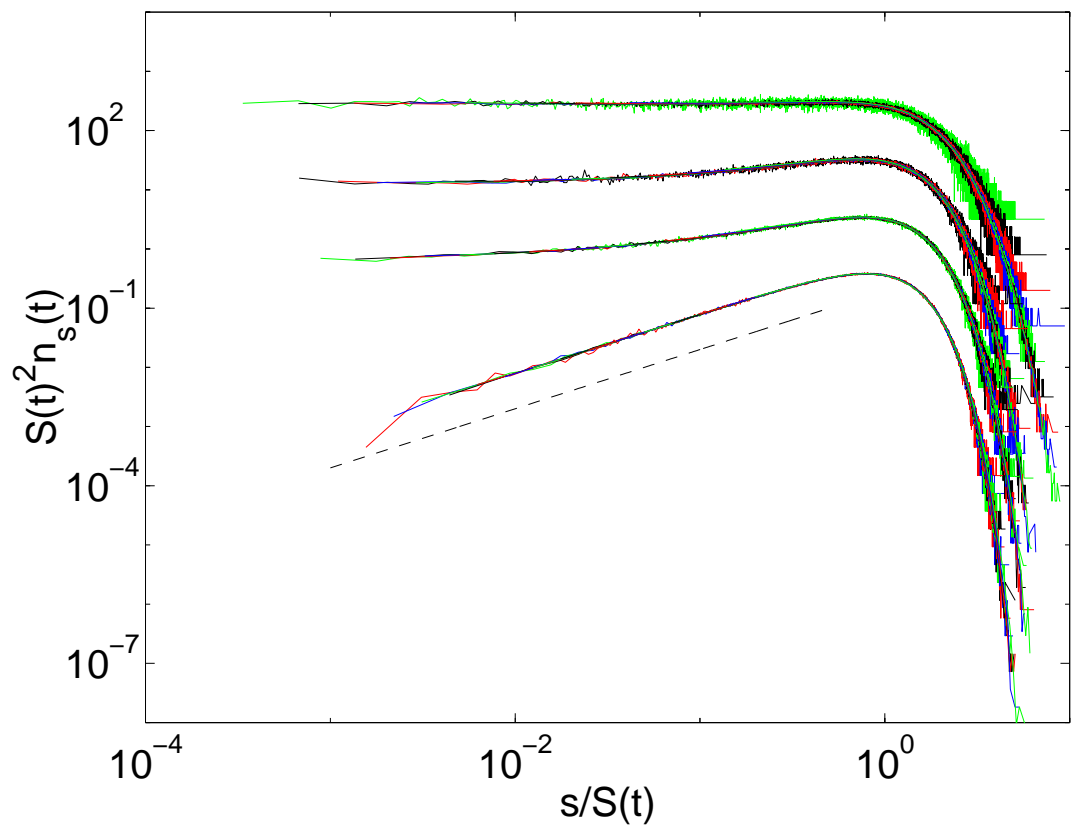}
\caption{The scaling of the cluster size distributions for $\gamma =$
0.00, 0.40, 0,57, and 1,00 (from bottom to top). Distributions are
shown at times from $t=2^3$ to $2^{k_{\rm max}}$, where $k_{\rm max} =$
17, 15, 14, and 11 for $\gamma =$ 0.00, 0.40, 0,57, and 1,00,
respectively. The dashed line has a slope 1.
The data are averaged over 50000 realizations on a
system of size $55555$. The distributions
(except the $\gamma = 0$ distribution)
have been shifted in the vertical direction for clarity.
}
\label{taufig}
\end{figure}

Simulations confirm this although
crossover effects make the analysis intractable near $\gamma =
0$. Figure~\ref{taufig} shows the scaling plots for the cluster size
distribution for various values of the diffusion exponent. The bigger
the $\gamma$ is the faster the scaling function approaches a constant
near $x=0$. For $\gamma \lesssim 0.5$ the times reached in simulations
are too small to reveal the asymptotic scaling behavior. In this
region also the measurements of $w$ from $n_s(t) \sim t^{-w}$ give
information only on the crossover effects~\cite{Hellen:PRE62}.

In summary, we have studied the survival of clusters in DLCA in one
dimension. 
The decay of the initial state equals the density
of clusters that stay intact by not aggregating with
others. This can be analyzed on the mean-field level
as the survival of a random walker bounded by two others with
time-dependent diffusion coefficients. This maps to diffusion in a
wedge with absorbing boundaries and time-dependent wedge angle.
For $\gamma>0$
the surviving particles are such that they pick the strategy
of staying immobile. The resulting survival exponent
is non-monotonic and discontinuous at $\gamma=0$. 
Cluster survival determines the polydispersity exponent
$\tau$ that characterizes the small cluster size tail. It is 
discontinuous and constant, $\tau=0$, for $0<\gamma<2$.  
For $\gamma<0$ the
survival probability decays stretched exponentially, the 
fluctuations of the neighboring clusters determine the stretching
exponent, and the mean-field RW-picture gives only a qualitative
understanding of the survival.

Above one dimension the study of cluster survival will be both
interesting and much less straightforward.
A similar mean-field picture in terms of
first-passage times of random walks is not directly applicable.
For example, one lacks much of the theory needed to
analyze the survival behavior of many interacting particles. We
conclude with the conjecture  
that the solution of the survival problem is related to
the cluster size distribution also in higher dimensions. 
There it should be possible to find experimental realizations,
to study the survival phenomenon \cite{expts}.

{\em Acknowledgements - }
We thank Erik Aurell and Amit K. Chattopadhyay for discussions.
This research has been
supported by the Academy of Finland's Center of Excellence program.

\end{document}